\documentclass[prb,aps,twocolumn,showpacs]{revtex4}%
\usepackage{graphicx}
\usepackage{dcolumn}
\usepackage{bm}
\usepackage{amsmath}%
\usepackage{amsfonts}%
\usepackage{amssymb}
\begin{document}
\title{Electron heating in metallic resistors at sub-Kelvin temperature}
\author {B.\ Huard}
\author {H.\ Pothier}
\author {D.\ Esteve}
\affiliation{Quantronics group, Service de Physique de l'Etat Condens\'{e} (CNRS URA 2464), DRECAM, CEA-Saclay, 91191
Gif-sur-Yvette, France}
\author {K. E. Nagaev}
\affiliation{Institute of Radioengineering and Electronics, Russian Academy of Sciences, Mokhovaya ulica 11, 125009
Moscow, Russia}
\date{\today     }

\begin{abstract}
In the presence of Joule heating, the electronic temperature in a metallic
resistor placed at sub-Kelvin temperatures can significantly exceed the phonon
temperature. Electron cooling proceeds mainly through two processes:
electronic diffusion to and from the connecting wires and electron-phonon
coupling.\ The goal of this paper is to present a general solution of the
problem, in a form that can easily be used in practical situations. As an
application, we compute two quantities that depend on the electronic
temperature profile: the second and the third cumulant of the current noise at
zero frequency, as a function of the voltage across the resistor. We also
consider time dependent heating, an issue relevant for experiments in which
current pulses are used, for instance in time-resolved calorimetry experiments.

\end{abstract}
\maketitle

\section{Motivations and outline}

When performing electrical measurements, the signal to noise ratio can usually
be improved by simply increasing the currents or voltages. In low-temperature
experiments, this procedure is problematic because of Joule heating, which can
affect the temperature of the circuit under investigation or the temperature
of resistors on bias lines, leading to excess noise. Particularly critical is
the situation in the sub-Kelvin range, because the temperature of the
electrons decouples from the lattice
temperature\cite{Arai,Roukes,Wellstood,PRB}. A very conservative, wide-spread
rule of thumb among experimentalists is that the voltage $V$ across the small
conductors should not exceed $k_{B}T_{\mathrm{ph}}/e,$ with $T_{\mathrm{ph}}$
the lattice (phonon) temperature. In contrast, macroscopic components, like
commercial resistors, are believed to be immune to electron heating. In fact,
the first rule is severe, and the second assertion is often incorrect. The
goal of the present article is to provide the experimental physicist with easy
evaluation tools of heating effects, in order to optimize experiments.

The important parameters are the voltage $V,$ the resistance $R,$ the lattice temperature $T_{\mathrm{ph}},$ the
resistor volume $\Omega$ and a parameter $\Sigma$ that describes electron-phonon coupling. The first step is to
calculate the characteristic temperature $T_{\Sigma}$, which is the temperature that the electrons would reach if
cooling would occur only through the coupling to a bath of zero-temperature phonons:
\begin{equation}
T_{\Sigma}=\left(  \frac{V^{2}}{\Sigma\Omega R}\right)  ^{1/5}. \label{TSigma}%
\end{equation}
The average electron temperature can then be directly read from Fig.\thinspace\ref{B}, the central result of this work,
in which the voltage $V$, the average temperature $T_{\mathrm{av}}$ and the lattice temperature $T_{\mathrm{ph}}$ are
all expressed in units of $T_{\Sigma}.$ In the section IIA, we explain how this result is obtained, and give analytical
expressions in various limits. The results are used to calculate the second and third cumulant of the current noise
produced by the resistor (section IIB, Fig.\thinspace\ref{C}). Numerical applications are carried out explicitely in
section IIC, showing in particular that heating in commercial resistors can be important. In section III, we address
time-dependent situations, and calculate how fast electrons heat up in a resistor when a current is applied, and how
fast they cool down when the current is switched off. For small voltages ($eV\ll k_{B}T_{\Sigma}$), the variations of
temperature in both transients is exponential, with the diffusion time $\tau_{D}$ across the whole conductor as a
characteristic time (see Fig.$\,$\ref{D}). In the opposite limit ($eV\gg k_{B}T_{\Sigma}$), heating is exponential, but
cooling proceeds very slowly, with a powerlaw dependence (see Fig.\thinspace\ref{E}).\ The timescale is the
electron-phonon scattering time $\tau_{\mathrm{e-ph}}$ at temperature $T_{\Sigma},$ defined by
Eq.\thinspace(\ref{taueph}). As a numerical application, we consider in section IIIC a situation where repeated current
pulses are applied to a resistor, and compute the time-dependence of the electron temperature (Fig.\thinspace\ref{F}).

\section{Stationary situations}

\subsection{\bigskip Solution of the heat equation}

\begin{figure}[b]
\begin{center}
\includegraphics[width=1\columnwidth,clip]{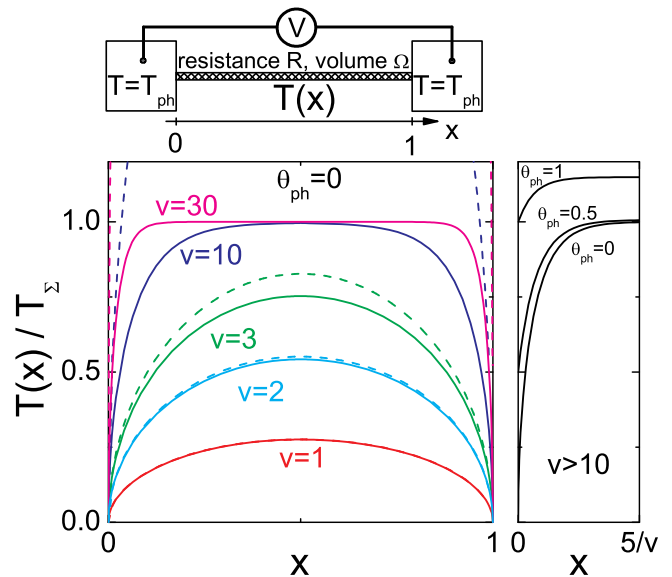}
\end{center}
\caption{(Color online) Top: Resistor biased by a voltage $V$ and placed
between two connecting wires in which the electron temperature $T$ and the
phonon temperature $T_{\mathrm{ph}}$ are equal. Left panel: Solid lines:
temperature profile in the resistor for different values of $v=eV/k_{B}%
T_{\Sigma}$ with $T_{\Sigma}=\left(  V^{2}/\Sigma\Omega R\right)  ^{1/5}$.
Dashed lines: temperature profile expected when phonon cooling is neglected
(Eq.\thinspace(\ref{hot})). Right panel: temperature profile near the ends of
the resistor for $v>10$ and $\theta_{\mathrm{ph}}=T_{\mathrm{ph}}/T_{\Sigma
}=0,$ 0.5, 1.}%
\label{A}%
\end{figure}

\begin{figure}[b]
\begin{center}
\includegraphics[width=1\columnwidth,clip]{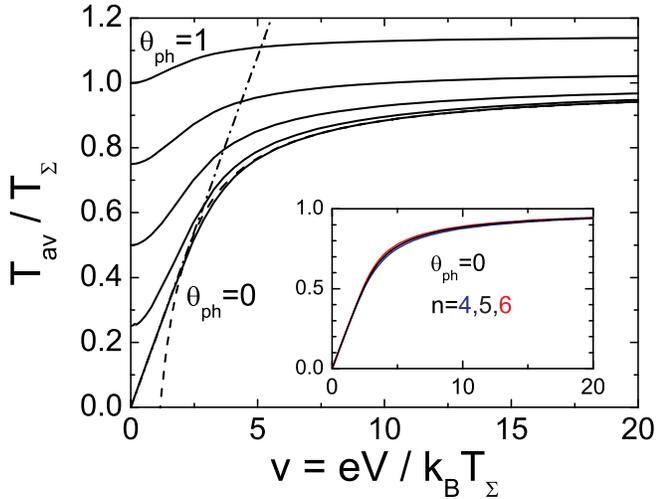}
\end{center}
\caption{(Color online) Average temperature $T_{av}$ in units of $T_{\Sigma},$ as a function of $v=eV/k_{B}T_{\Sigma},$
for $\theta_{ph}=T_{ph}/T_{\Sigma}=0,$ 0.25, 0.5, 0.75, 1 (from bottom to top). The dotted line corresponds to the
low-$v$ approximation $T_{av}=\frac{\sqrt{3}}{8}\frac{eV}{k_{B}}$, the dashed-dotted line to
the large-$v$ approximation $T_{av}=1-1.16/v.$ Inset: at $\theta_{ph}%
=T_{ph}/T_{\Sigma}=0,$ comparison of the evolution of the average temperature
with $v$ for various exponents $n$ of the temperature in the expression of the
heat flow through electron-phonon coupling.}%
\label{B}%
\end{figure}

When a voltage $V$ is applied to a two-terminal resistor (see top of Fig.\thinspace\ref{A}), the Joule power $V^{2}/R$
is delivered to the electrons. This power can dissipate by two mechanisms: the first one, which dominates at room
temperature or in macroscopic resistors, is phonon emission. It follows, at temperatures well below the Debye
temperature, a $T^{n}(x)-T_{\mathrm{ph}}^{n}$ dependence, with $4\leq n\leq6,$ and $T(x)$ the local electron
temperature, $T_{\mathrm{ph}}$ the phonon temperature\cite{Sergeev}. The second mechanism is the simple diffusion of
the energetic electrons out of the resistor. The energy is then dissipated in the connecting leads, which are, in
typical situations, large and low-resistive. The balance between the Joule power and the two cooling mechanisms can be
expressed in the form of a heat equation\cite{Steinbach}%
\begin{equation}
\frac{d}{dx}\left(  \frac{L_{o}T(x)}{R}\frac{d}{dx}T(x)\right)  =-\frac{V^{2}%
}{R}+\Sigma\Omega\left(  T^{5}(x)-T_{\mathrm{ph}}^{5}\right)  \label{heatbrut}%
\end{equation}
with $x$ the position along the resistor in reduced units ($x$ runs from 0 to 1), $L_{o}=\pi^{2}k_{B}^{2}/3e^{2}$ the
Lorenz number, $\Omega$ the resistor volume, $\Sigma$ the electron-phonon coupling constant (typically
$\Sigma\simeq2~\mathrm{nW}/\mathrm{\mu m}^{3}/\mathrm{K}^{5}$ for good metals \cite{Giazotto}). The left hand side of
Eq.\thinspace(\ref{heatbrut}) accounts for heat transport by electron diffusion, which is expressed by the
Wiedemann-Franz law, stating that the electron thermal conductivity is proportional to the product of the electrical
conductivity and the electron temperature. The following assumptions have been made to write Eq.\thinspace
(\ref{heatbrut}):

\begin{enumerate}
\item The electron temperature $T(x)$ is assumed to be well defined locally,
\textit{i.e. }the local electron energy distribution function is a Fermi
function. This requires that the thermalization of electrons among themselves
(e.g. by Coulomb interaction) occurs faster than the diffusion of electrons
across the resistor\cite{pothier}, a condition usually obeyed except for short
wires (length $\lesssim50\,\mathrm{%
%TCIMACRO{\U{b5}}%
%BeginExpansion
\mu
%EndExpansion
m}$) made of very pure materials\cite{Huard}.

\item The last term of the equation, which describes cooling by phonons,
assumes that the lattice temperature $T_{\mathrm{ph}}$ does not depend on the
local electron temperature $T(x).$ Corrections due to the Kapitza resistance
between the phonons of the resistive film and the substrate could in principle
be included\cite{Swartz,Wellstood}, but their contribution is not essential in practice.

\item The heat power transfered to phonons was taken proportional to $T^{n}(x)-T_{\mathrm{ph}}^{n}$ with $n=5.$
Theoretically, it is predicted that the exponent $n,$ which is related to a $E^{2-n}$ dependence of the electron-phonon
scattering rate with electron energy $E,$ can range from 4 to 6, depending on the relative sizes of the thermal phonon
wavelength and the electron mean free path, on the dimensionality of the phonon system, and on the dynamics of
impurities\cite{Sergeev}. In most experiments, values close to $n=5$ have been found (see discussions in
Refs.\thinspace \onlinecite  {Schmidt,Sergeev,Giazotto}), therefore our choice. The calculations can however be easily
extended to other values of $n,$ and the results presented in Fig.\thinspace\ref{B} apply, with another definition of
$T_{\Sigma},$ as discussed in the following.

\item Radiative cooling\cite{Schmidt}, which has a negligible effect in
resistors connected to large, non-superconducting contacts\cite{guichard}, is neglected.
\end{enumerate}

The heat equation (\ref{heatbrut}) has to be solved with boundary conditions
for $T(x)$ at $x=0$ and $x=1.$ When the connecting wires to the resistor are
low-resistive and very large compared to the resistor, as is the case for
macroscopic resistors made of thin and narrow metallic stripes of metal, one
can assume $T(0)=T(1)=T_{\mathrm{ph}}$. This simple hypothesis will be made in
the following. For on-chip thin-film resistors, heating of the contact pads
themselves may however not be negligible\cite{Henny}.

Before a general solution of Eq.\thinspace(\ref{heatbrut}) is presented, we
recall simple limits. The so-called interacting hot-electron
limit\cite{Steinbach} is obtained by neglecting phonon cooling:%
\begin{equation}
T(x)=\sqrt{T_{\mathrm{ph}}^{2}+\frac{3}{\pi^{2}}x(1-x)\left(  \frac{eV}{k_{B}%
}\right)  ^{2}} \label{hot}%
\end{equation}
(see left panel of Fig.\thinspace\ref{A}, dashed lines). For $T_{\mathrm{ph}%
}=0,$ the maximal temperature is $T(\frac{1}{2})=(\sqrt{3}/2\pi)\,eV/k_{B}%
\simeq0.28\,eV/k_{B},$ and the average temperature is $T_{\mathrm{av}}%
=\int_{0}^{1}T(x)dx=(\sqrt{3}/8)\,eV/k_{B}\simeq0.22\,eV/k_{B}.$
Electron-phonon coupling further reduces the temperature, so that this is an
upper bound on the average electron temperature, which numerically reads,
keeping\cite{noteTav} now $T_{\mathrm{ph}}$:%
\begin{equation}
\frac{T_{\mathrm{av}}}{T_{\mathrm{ph}}}\leq\sqrt{1+\left(  0.22\frac{eV}%
{k_{B}T_{\mathrm{ph}}}\right)  ^{2}}.
\end{equation}
In particular, the rule of thumb $eV=k_{B}T_{\mathrm{ph}}$ corresponds to a
2.5\% average overheating of the electrons$.$ Numerically, one obtains the
equivalent expression%
\begin{equation}
T_{\mathrm{av}\,}\text{[mK]}\leq\sqrt{T_{\mathrm{ph}}^{2}\text{[mK]+}%
(2.5\times V\,\text{[}\mathrm{%
%TCIMACRO{\U{b5}}%
%BeginExpansion
\mu
%EndExpansion
V])}^{2}}. \label{average T}%
\end{equation}

\begin{figure}[bb]
\begin{center}
\includegraphics[width=1\columnwidth,clip]{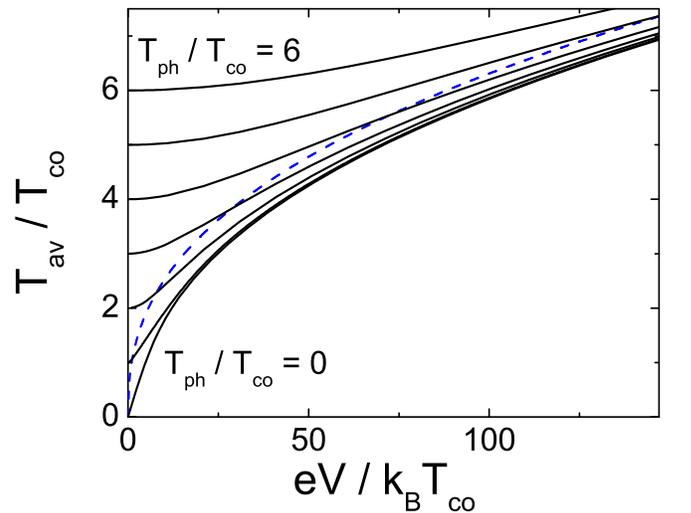}
\end{center}
\caption{(Color online) Average temperature $T_{av}$ as a function of voltage
$V,$ for various temperatures $T_{\mathrm{ph}},$ all in units of
$T_{co}=\left(  \Sigma\Omega Re^{2}/k_{B}^{2}\right)  ^{-1/3}.$ The value of
$T_{\mathrm{ph}}/T_{co}$ is given by the intersection of the curves with the
vertical axis. Blue dashed line is the reference temperature $T_{\Sigma}.$}%
\label{B2}%
\end{figure}

In the opposite limit where cooling by diffusion can be neglected, the
electron temperature is homogeneous and equal to\cite{Wellstood}
\begin{equation}
T=\left(  T_{\mathrm{ph}}^{5}+T_{\Sigma}^{5}\right)  ^{1/5}.
\label{phonons only}%
\end{equation}
In the following, we call this limit ``the fully thermalized regime''. In the limit $T_{\mathrm{ph}}=0,$ the
temperature grows as $V^{2/5}$ (see Fig.$\,$\ref{B2})$.$

In intermediate regimes, the temperature profile is obtained by solving
numerically the heat equation. For generality, it is convenient to rewrite it
in reduced units. One possibility is to take as a reference the ``cross-over
temperature''\cite{PRB} $T_{\mathrm{co}}=\left(  \Sigma\Omega Re^{2}/k_{B}%
^{2}\right)  ^{-1/3}$, which is the energy scale for which phonon cooling and
diffusion cooling are equaly important. This temperature is an intrinsic
quantity for the resistor, which in particular does not depend on $V$ or
$T_{\mathrm{ph}}.$ However, it does not correspond to the electron temperature
in any limit, therefore we prefer to take as a reference $T_{\Sigma}$, keeping
in mind that it depends on $V.$ Defining $\theta(x)=T(x)/T_{\Sigma},$
$v=eV/k_{B}T_{\Sigma}$ ($=(eV/k_{B}T_{\mathrm{co}})^{3/5}$), $\theta
_{\mathrm{ph}}=T_{\mathrm{ph}}/T_{\Sigma}$ ($\propto V^{-2/5}$),
Eq.\thinspace(\ref{heatbrut}) reads:%
\begin{equation}
\frac{d^{2}}{dx^{2}}\theta^{2}(x)=\frac{6}{\pi^{2}}v^{2}\left(  \theta
^{5}(x)-\theta_{\mathrm{ph}}^{5}-1\right)  . \label{eqchaleur}%
\end{equation}
The temperature profile being symmetric with respect to the middle of the
wire, Eq.\thinspace(\ref{eqchaleur}) needs to be solved for $0<x<\frac{1}{2}$
with the boundary conditions $\theta(0)=\theta_{\mathrm{ph}}$ and
$\theta^{\prime}(\frac{1}{2})=0.$ Instead of solving this non-linear
differential equation with boundary conditions specified at different points,
it is convenient to rewrite it in the following integral form:%
\begin{equation}
\frac{2\sqrt{3}v}{\pi\theta_{m}}x=\int_{\theta_{\mathrm{ph}}^{2}/\theta
_{m}^{2}}^{\theta^{2}(x)/\theta_{m}^{2}}\mathrm{d}u\,\left(  \frac{2}{7}%
\theta_{m}^{5}\left(  u^{7/2}-1\right)  -\lambda^{5}(u-1)\right)  ^{-1/2}
\label{eq chaleur reduite}%
\end{equation}
with $\theta_{m}=\theta(\frac{1}{2})$ and $\lambda=\left(  \theta _{\mathrm{ph}}^{5}+1\right)  ^{1/5}.$ The value of
$\theta_{m}$ is obtained by solving Eq.\thinspace(\ref{eq chaleur reduite}) for $x=1/2.$ In the left panel of
Fig.\thinspace\ref{A}, the temperature profile along the resistor is given for $v=1,2,3,10$ and 30, assuming
$T_{\mathrm{ph}}=0$.\ At $v\lesssim1,$ one recovers the result of Eq.\thinspace(\ref{hot}), plotted as dashed lines:
phonon cooling can be neglected. For $v\gtrsim10$, it is an excellent approximation to take $\theta_{m}=\lambda$ in
Eq.\thinspace(\ref{eq chaleur reduite}), a value that does not depend on $v$. $T(x)$ is then a function of $xv$ only,
which is essentially constant around the middle of the wire, whereas at a distance $5/v$ from the contacts one obtains
the profiles shown in the right panel of Fig.\thinspace\ref{A} for $\theta_{\mathrm{ph}}=0$ (see note
\onlinecite{approx})$,$ 0.5 and 1. The characteristic length over which the electron temperature varies from
$T_{\mathrm{ph}}$ to $T_{\Sigma}$ is therefore $L_{\Sigma}=L/v=\frac{\sqrt{3}}{8}L^{3/5}L_{\mathrm{e-ph}}^{2/5},$
where the ``electron-phonon length''%
\begin{equation}
L_{\mathrm{e-ph}}=\left(  \frac{8k_{B}}{\sqrt{3}e}\right)  ^{5/2}\left(
\rho\Sigma V^{3}\right)  ^{-1/2} \label{Leph}%
\end{equation}
($\rho$ being the resistivity) is defined, following\cite{errorS} Ref.\thinspace\onlinecite{Steinbach}, as the resistor
length for which $4k_{B}T_{\Sigma}/R$ is equal to the current noise in the interacting hot-electron regime
$\frac{\sqrt{3}}{2}eI.$ This length is typically of the order of a few
%TCIMACRO{\U{b5}}%
%BeginExpansion
$\mu$%
%EndExpansion
m for voltages of the order ot 1\thinspace mV.



From the complete temperature profile $\theta(x),$ the average electron temperature $T_{\mathrm{av}}$ is obtained using
$T_{\mathrm{av}}=T_{\Sigma }\int_{0}^{1}\theta(x)\,\mathrm{d}x.$ The central result of this work is the resulting plot,
shown in Fig.~\ref{B}, of the average temperature $T_{\mathrm{av}}$ as a function of the voltage $V$, both in units of
$T_{\Sigma},$ for $T_{\mathrm{ph}}/T_{\Sigma}=0,$ 0.25, 0.5, 0.75 and 1. At
$\theta_{\mathrm{ph}}=0$ and $v\lesssim2,$ $T_{\mathrm{av}}\sim\frac{\sqrt{3}%
}{8}\frac{eV}{k_{B}}$ (dotted line), whereas for $v\gtrsim4,$ $T_{\mathrm{av}%
}/T_{\Sigma}\approx1-1.16/v$ (dashed line). This $1/v$ dependence is due to the crossover regions of width $\sim5/v$ at
the resistor ends. Figure~\ref{B} can
be directly used to read out the average electron temperature $T_{\mathrm{av}%
}$ for a given set of experimental parameters $(V,$ $T_{\mathrm{ph}}),$ after
having computed $T_{\Sigma}$ with Eq.\thinspace(\ref{TSigma}). Interestingly,
the corresponding curves for other exponents of the temperature in the last
term of Eq.\thinspace(\ref{heatbrut}) ($n=4$ or 6 instead of 5) are almost
identical (see inset), and the same curves can be used to evaluate
$T_{\mathrm{av}},$ however with the generalized definition of the reference
temperature $T_{\Sigma}=$ $\left(  V^{2}/\Sigma\Omega R\right)  ^{1/n}.$

However, because of the use of reduced units which depend on $V,$ the
$v$-dependence of $T_{\mathrm{av}}$ at a fixed value of $\theta_{\mathrm{ph}}$
shown in Fig.\thinspace\ref{B} does \textit{not} correspond to a situation in
which $V$ is changed at a fixed $T_{\mathrm{ph}},$ since $\theta_{\mathrm{ph}%
}\propto V^{-2/5}.$ In order to visualize how temperature increases with $V$
at a given phonon temperature, we plot in Fig.\thinspace\ref{B2} the average
temperature $T_{\mathrm{av}}(V)$ for various $T_{\mathrm{ph}},$ with $V$ and
$T_{\mathrm{ph}}$ given in units of\cite{PRB} $T_{\mathrm{co}}=\left(
\Sigma\Omega Re^{2}/k_{B}^{2}\right)  ^{-1/3},$ which is constant for a given
resistor. The range in voltage $V$ is the same as in Fig.\thinspace\ref{B}. We
have used the relations $v^{5/3}=eV/kT_{\mathrm{co}}$; $\left(  T_{\mathrm{av}%
}/T_{\Sigma}\right)  v^{2/3}=T_{\mathrm{av}}/T_{\mathrm{co}}.$

\subsection{\bigskip Second and third cumulants of the current noise at zero frequency}

The temperature profile can be used to evaluate the current noise properties
of the resistor. We focus here on the second ($S_{2}$) and third ($S_{3}$)
cumulants of noise at low frequencies ($\hslash\omega\ll eV$ for $S_{2},$
$\hslash\omega\ll eV,\hbar/\tau_{D}$ for $S_{3}$ (see Ref.\thinspace
\onlinecite{Pilgram}  ), with $\tau_{D}=L^{2}/D$ the diffusion time, $D$ the
diffusion constant):
\[
S_{2}=2\iint\mathrm{d}t\,\left\langle \delta I(0)\delta I(t)\right\rangle
\]
and
\[
S_{3}=\iint\mathrm{d}t_{1}\mathrm{d}t_{2}\,\left\langle \delta I(0)\delta
I(t_{1})\delta I(t_{2})\right\rangle
\]
with $\delta I(t)=I(t)-\left\langle I\right\rangle .$ It has been shown that
when phonon cooling can be disregarded and $eV\gg k_{B}T_{\mathrm{ph}},$
$S_{2}$ and $S_{3}$ are proportional to the applied current
\cite{Steinbach,Henny}: $S_{2}=F_{2}\times2eI$ and $S_{3}=F_{3}\times e^{2}I$
with $F_{2}$ and $F_{3}$ generalized ``Fano factors''. When furthermore the
rate of electron-electron interaction is negligible compared to $1/\tau_{D}$,
the distribution function is not a Fermi function, but a function with two
steps \cite{Nagaev,pothier}, and\cite{Been1,Nagaev,Lee} $F_{2}=\frac{1}%
{3}\approx0.33;$ $F_{3}=\frac{1}{15}\approx0.067.$ In the opposite limit,
where electron-electron interaction is strong, electrons thermalize locally to
distribute in energy according a Fermi function, and the temperature profile
is given by Eq.\thinspace(\ref{hot}). One then obtains\cite{NagaevN,Kozub}
$F_{2}=\frac{\sqrt{3}}{4}\approx0.43$ and\cite{Gutman} $F_{3}=\frac{8}{\pi
^{2}}-\frac{9}{16}\approx0.248.$

In presence of strong phonon cooling ($v\gg1$), the electron temperature
becomes homogeneous, at a value $T_{\Sigma}$ smaller than $eV/k_{B}.$ It is
then expected\cite{Gutman} that $F_{2},F_{3}\rightarrow0.$

For intermediate coupling to phonons, $F_{2}$ and $F_{3}$ depend on the
voltage across the resistor. Their value are obtained from the full solution
of the heat equation: the second cumulant is given by a Johnson-Nyquist-like
formula\cite{Steinbach} $S_{2}=4k_{B}T_{\mathrm{av}}/R$ in which the noise
temperature is the average electron temperature $T_{\mathrm{av}},$ yielding
$F_{2}=2k_{B}T_{\mathrm{av}}/eV=2\theta_{\mathrm{av}}/v$. This formula can be
understood as resulting from the added Johnson-Nyquist noise of small sections
of the resistor, each at a temperature $T(x).$ The decay of $F_{2}$ at large
$V$ was discussed in Ref.\thinspace\thinspace\onlinecite{NagaevN}  , and the
complete crossover was calculated in Ref.\thinspace\thinspace
\onlinecite{Naveh}  by numerical integration of Eq.\thinspace(\ref{heatbrut}).
In turn, $F_{3}$ is given by (see Appendix)%
\begin{equation}
\bigskip F_{3}=\frac{36}{\pi^{2}}\int\nolimits_{0}^{1}\mathrm{d}%
x\,\mathrm{d}y\frac{1}{\theta(x)}\,G_{1}(\theta,x,y)\left\{  \theta
(y)-2\theta_{\mathrm{av}}\right\}
\end{equation}
where $G_{1}(\theta,x,y)$ the Green's function such that $\left(  \nabla
^{2}+\frac{15}{\pi^{2}}\,v^{2}\,\theta^{3}(x)\right)  G_{1}(\theta
,x,y)=\delta(x-y)$ and $G_{1}(\theta,0,y)=G_{1}(\theta,x,0)=0.$ The
calculation of $F_{3}$ is detailed in the Appendix. The right panel of
Fig.\thinspace\ref{C} shows the voltage dependence of $F_{2}$ (blue line) and
$F_{3}$ (red line) as a function of $v$ (bottom axis) and $L/L_{\mathrm{e-ph}%
}=\left(  \sqrt{3}v/8\right)  ^{5/2}$ (top axis), for $T_{\mathrm{ph}}=0$. Also shown with a dashed line is the curve
obtained for $F_{2}$ when electron diffusion is neglected\cite{NagaevN}, using Eq.\thinspace(\ref{phonons only})
(dashed line), which gives $F_{2}=2/v\propto V^{-3/5}$. In turn, at large voltages, $F_{3}\propto v^{-2}\propto
V^{-6/5}.$ If one considers a situation where the resistor length $L$ is varied at constant current, then
$F_{2}\propto1/L$ and $F_{3}\propto1/L^{2}.$ The decay to zero of $F_{2}$ and $F_{3}$ from the interacting hot-electron
values ($\frac{\sqrt{3}}{4}$ and $\frac{8}{\pi^{2}}-\frac{9}{16}$) is therefore very slow, as already pointed in
Ref.\thinspace\onlinecite{Naveh,Gutman}. The non-Gaussian character of the current noise, evidenced by $F_{3}\neq0,$ is
washed out at $L/L_{\mathrm{e-ph}}\gtrsim10.$

\begin{figure}[b]
\begin{center}
\includegraphics[width=1\columnwidth,clip]{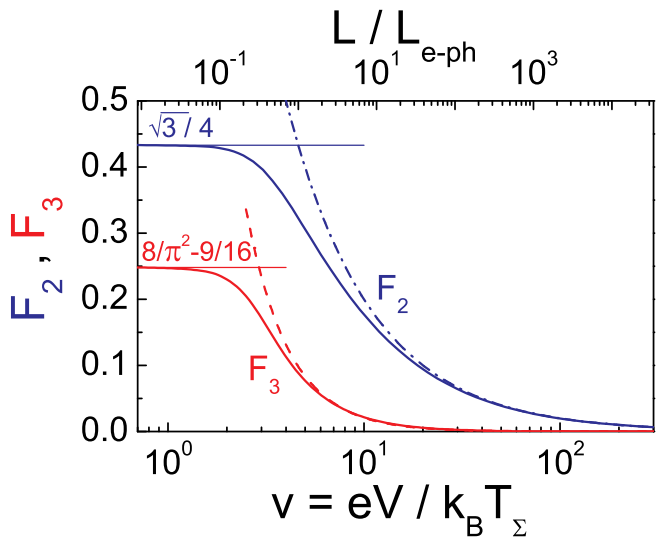}
\end{center}
\caption{(Color online) Solid lines: Fano factors $F_{2}=S_{2}/2eI$ (blue) and $F_{3}=S_{3}/e^{2}I$ (red) for the
zero-frequency second and third cumulant of noise, from the hot electron limit to the fully thermalized regime.
Dashed-dotted line: asymptotic dependence of $F_{2}$ neglecting electron diffusion: $F_{2}=2/v.$ Top axis is resistor
length over $L_{\mathrm{e-ph}}$ (see Eq.\thinspace
(\ref{Leph})).}%
\label{C}%
\end{figure}

\subsection{Examples}

We consider here a few cases illustrating the use of the results given in the
preceding sections. As a first example, we consider a $10\,\mu\mathrm{m}%
$-long, $100\,\mathrm{nm}$-wide and $5\,\mathrm{nm}$-thick Cr resistor with resistance $R=25\,\mathrm{k}\Omega$ like
these used in Ref.\thinspace \onlinecite{Joyez}, biased at $V=1\,\mathrm{mV}$ and placed at
$T_{\mathrm{ph}}=20\,\mathrm{mK.}$ Assuming $\Sigma_{\mathrm{Cr}%
}=2\,\mathrm{nW}/\mathrm{\mu m}^{3}/\mathrm{K}^{5},$ the characteristic
temperature is $T_{\Sigma}=1.3\,$K, and the voltage and phonon temperature in
reduced units $v=eV/k_{B}T_{\Sigma}\sim9$ and $\theta_{\mathrm{ph}}=0.015.$
The noise temperature is directly read from Fig.\thinspace\ref{B}:
$T_{\mathrm{av}}\sim0.87\times T_{\Sigma}\sim1.15\,$K. At this voltage,
heating of the resistor is thus very important, an effect which hindered the
authors of Ref.\thinspace\onlinecite{Joyez}  from drawing clear-cut
conclusions from Coulomb blockade measurements at finite voltage. Increasing
the resistor volume $\Omega$ with ``cooling fins'' can help decreasing
electron heating \cite{Wellstood}, but such a procedure is extremely
inefficient since the characteristic temperature $T_{\Sigma}$ decreases as
$\Omega^{-1/5}$ only.

As a second application, we now consider a commercial macroscopic surface
mount resistor, with $R=500\,\,\Omega.$ Such resistors, made of thin
($\sim10\,\mathrm{nm}$) NiCr films\cite{theseBH} with resistivity $\rho
\sim100\,\mu\Omega\mathrm{\,cm}$ and dimensions $\sim$1\thinspace mm$\times
$0.2\thinspace mm, were used as bias resistors in measurements of the state of
superconducting Josephson Q-bits\cite{Vion,theseAC} performed at
$15\,\mathrm{mK}$, with a bias current $\sim0.8\,\mu\mathrm{A}$, resulting in
a voltage $V\sim400\,\mu\mathrm{V}.$ The corresponding temperature scale
$T_{\Sigma}\sim150\,\mathrm{mK}$ yields $v=eV/k_{B}T_{\Sigma}\sim30$ and
$\theta_{\mathrm{ph}}\sim0.1,$ hence, from Fig.\thinspace\ref{B},
$T_{\mathrm{av}}\sim T_{\Sigma}\sim150\,\mathrm{mK.}$ Even in such a
macroscopic resistor, the volume is not sufficient to provide with enough
electron-phonon coupling, and heating is important. In the next section, we
show how this heating is limited when pulses are used instead of static voltages.

\section{Time-dependent situations: switching on and off Joule heating}

The case of a constant voltage $V$ across the resistor, which was investigated
above, can be extended to the case of slowly varying voltages directly.
However, when $V$ changes on timescales shorter than the diffusion time or
than the electron-phonon scattering time\cite{time-dep} (see below), the
previous results cannot be used to calculate instantaneous temperatures. These
issues are solved by adding to the heat equation (\ref{heatbrut}) a
time-dependent term $dQ/dt=C_{e}dT/dt,$ with $C_{e}=\gamma T\Omega$ the
electronic heat capacity, $\gamma=(\pi^{2}/3)k_{B}^{2}\nu_{F}$ (from Fermi
liquid theory), with $\nu_{F}$ the density of states at Fermi energy (spin
degeneracy included). When $V(t)=Vf(t),$ the time-dependent heat equation can
then be rewritten, in reduced units, as
\begin{equation}
\frac{\partial\theta^{2}}{\partial\tau}=\frac{\partial^{2}\theta^{2}}{\partial
x^{2}}-\frac{6}{\pi^{2}}v^{2}\left(  \theta^{5}-\theta_{\mathrm{ph}}^{5}%
-f^{2}(\tau)\right)  \label{heat equation with time}%
\end{equation}
where $\tau=t/\tau_{D}$ is the reduced time with $\tau_{D}=L^{2}/D$ the
diffusion time. Note that the reference temperature $T_{\Sigma}$ used to
define $\theta(x)=T(x)/T_{\Sigma}$ is calculated with the voltage scale $V,$
not with the time-dependent value $V(t).$ In the following, we treat more
explicitely two situations: how a resistor heats up when the voltage is
applied at $t=0,$ \textit{i.e.} $f(\tau)=H(\tau),$ and how a resitor cools
down when the voltage is set to zero at $t=0,$ \textit{i.e.} $f(\tau
)=1-H(\tau)$. Here, $H(\tau)$ is the Heaviside function ($0$ for $\tau<0,$ 1
for $\tau>0)$. These situations also allow to describe experiments in which
current or voltage pulses are used like, for example, when measuring the
switching rate of Josephson junctions\cite{Vion}. Understanding how the pulse
characteristics can reduce the noise in such measurements is therefore
important to design the readout of superconducting Q-bits.

When a voltage is applied, the linear drop of the electrical potential, which
results from the collective charge modes, establishes after an $RC$ time,
where the capacitance $C$ is the capacitance of the wire to ground. This time
is generally much shorter than the time necessary to build up the temperature
profile, which involves diffusion of individual electrons. Hence, we consider
here that Joule heating is homogeneous as soon as a voltage is applied. When
$v\lesssim1,$ the temperature profile is entirely determined by the
temperature at the ends of the resistor, therefore a steady-state regime is
reached only when the electrons have diffused across the whole resistor and
the characteristic time is the diffusion time $\tau_{D}$. If $v\gg1,$ the
transient is shorter because, apart from very close to the ends, the
temperature is mostly determined by a local equilibrium between Joule heating
and phonon emission. We now treat quantitatively these two limits$.$

\subsection{Small $v$ limit}

\begin{figure}[b]
\begin{center}
\includegraphics[width=1\columnwidth,clip]{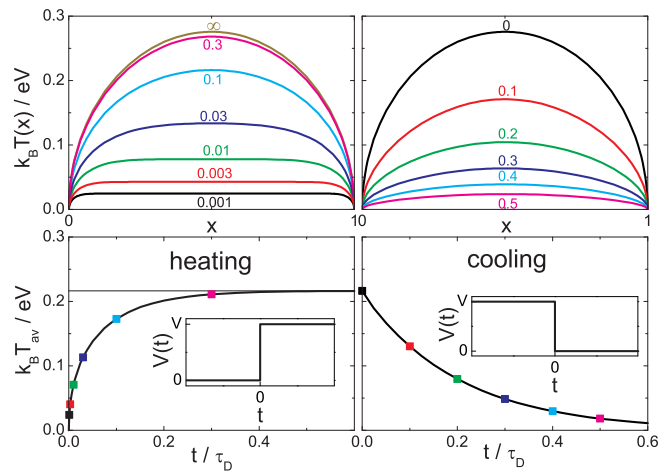}
\end{center}
\caption{(Color online) Time evolution of the temperature profile (top panels)
and of the average temperature $T_{av}$ (bottom panels) in the limit $v\ll1,$
for $T_{ph}=0.$ Left panels: heating sequence of the resistor when
$V(t)/V=H(t)$ (as shown in inset); right panels: cooling sequence of the
resistor when $V(t)/V=1-H(t)$ (as shown in inset). The profiles are plotted at
various values of $t/\tau_{D},$ with $\tau_{D}=L^{2}/D$ the diffusion time.
The colors of the solid curves in the top panels corresponds to those of the
square symbols in the bottom panels. The curves in the bottom panels cannot be
distinguished from $k_{B}T_{av}/eV\simeq\frac{\sqrt{3}}{8}\sqrt{1-\exp
(-10\tau)}$ and $\frac{\sqrt{3}}{8}\exp(-5\tau),$ respectively.}%
\label{D}%
\end{figure}

If $v\lesssim1,$ $\theta(x,t)-\theta_{\mathrm{ph}}\ll1$ even when the
stationary regime is reached, and Eq.\thinspace(\ref{heat equation with time})
reduces to
\begin{equation}
\frac{\partial\theta^{2}}{\partial\tau}=\frac{\partial^{2}\theta^{2}}{\partial
x^{2}}+\frac{6}{\pi^{2}}v^{2}f^{2}(\tau). \label{timedep}%
\end{equation}
As shown in section IIA, the proper energy scale when $v\lesssim1$ is $eV,$
and the solution of Eq.\thinspace(\ref{timedep}) that satisfies the boundary
conditions $T(0,\tau)=T(1,\tau)=T_{\mathrm{ph}}$ reads%
\begin{equation}
\left(  \frac{k_{B}T(x,\tau)}{eV}\right)  ^{2}=\left(  \frac{k_{B}%
T_{\mathrm{ph}}}{eV}\right)  ^{2}+\sum\limits_{k\text{ odd}}a_{k}(\tau
)\sin(\pi kx) \label{Tdexett}%
\end{equation}
\ with $a_{k}(\tau)$ solution of
\begin{equation}
\frac{\mathrm{d}a_{k}(\tau)}{\mathrm{d}\tau}+\pi^{2}k^{2}a_{k}(\tau
)=\frac{24}{\pi^{3}k}f^{2}(\tau).
\end{equation}
In the case where $f(\tau)=H(\tau),$ heating is then given by $a_{k}%
(\tau)=\left(  24/\pi^{5}k^{3}\right)  (1-e^{-\pi^{2}k^{2}\tau}),$ and in the
case $f(\tau)=1-H(\tau),$ cooling from the profile (\ref{hot}) follows
$a_{k}(\tau)=\left(  24/\pi^{5}k^{3}\right)  e^{-\pi^{2}k^{2}\tau}.$
Corresponding temperature profiles at various times are plotted in the top
panels of Fig.\thinspace\ref{D} assuming $T_{\mathrm{ph}}=0$, whereas the time
evolution of the average temperature $T_{\mathrm{av}}$ is plotted in the
bottom panels. At very short times, the average temperature grows as
$k_{B}T_{\mathrm{av}}/eV=\sqrt{6\,\tau}/\pi.$ At $\tau\gtrsim0.01,$ a better
approximation is
\begin{equation}
\frac{k_{B}T_{\mathrm{av}}}{eV}\simeq\frac{\sqrt{3}}{8}\sqrt{1-\exp
(-10\,\tau)},
\end{equation}
which cannot be distinguished from the exact solution in Fig.$\,$\ref{D}$.$ At
$\tau\simeq0.5,$ the asymptotical temperature profile given by Eq.\thinspace
(\ref{heatbrut}) is essentially established. Conversely, after the voltage is
turned off, the temperature decay is well approximated (within the line width
in Fig.$\,$\ref{D}) by%
\begin{equation}
\frac{k_{B}T_{\mathrm{av}}}{eV}\simeq\frac{\sqrt{3}}{8}\exp(-5\,\tau).
\label{cooldiff}%
\end{equation}
Hence, in the low-voltage regime, heating and cooling occur exponentially, and
the timescale is the diffusion time $\tau_{D}.$ In the example of the
$500\,\,\Omega$ commercial resistor in section IIB, $\tau_{D}\sim
3\,\mathrm{ms.}$ For metallic wires made of pure materials\cite{theseBH} with
an elastic mean free path of the order of 40\thinspace cm$^{2}$/s, $\tau
_{D}\sim20\,\mathrm{ns}$ for a length $L\sim20\,\mathrm{%
%TCIMACRO{\U{b5}}%
%BeginExpansion
\mu
%EndExpansion
m.}$

\subsection{Large $v$ limit}

\begin{figure}[b]
\begin{center}
\includegraphics[width=1\columnwidth,clip]{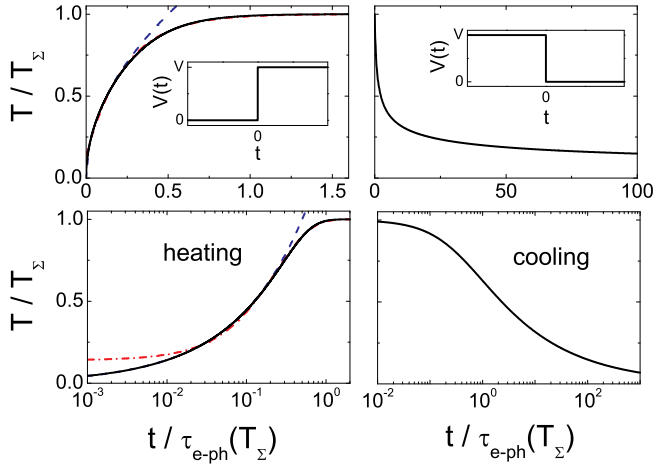}
\end{center}
\caption{(Color online) Evolution of the temperature with time in linear (top) and log (bottom) scale in the limit
$v\gg1,$ for $T_{ph}=0.$ Left panels: heating sequence of the resistor when $V(t)/V=H(t)$ (as shown in inset); right
panels: cooling sequence of the resistor when $V(t)/V=1-H(t)$ (as shown in inset)$.$ Times are given in units of the
electron-phonon time at temperature $T_{\Sigma}:$ $\tau_{e-ph}\left(  T_{\Sigma}\right)  =\gamma/\left(  \Sigma
T_{\Sigma}^{3}\right) .$ Blue dashed line is $\sqrt{2\,\tau^{\ast}}$, red dashed-dotted line is $1-0.86\exp(-4.2\,\tau^{\ast}),$ with $\tau^{\ast}%
=t/\tau_{e-ph}\left(  T_{\Sigma}\right)  .$}%
\label{E}%
\end{figure}

If $v\gg1,$ it was shown in section IIA that the temperature becomes almost
homogeneous in the wire. The relevant timescale is then the electron-phonon
scattering time\cite{Schmidt,factor,notetaueph} at the characteristic
temperature $T_{\Sigma}:$
\begin{equation}
\tau_{\mathrm{e-ph}}\left(  T_{\Sigma}\right)  =\frac{\gamma}{\Sigma
T_{\Sigma}^{3}}=\frac{\pi^{2}}{3}\frac{L_{\Sigma}^{2}}{D} \label{taueph}%
\end{equation}
with $L_{\Sigma}$ the characteristic length for the variation of $T(x)$
introduced in section IIA. Numerically, $\gamma/\Sigma\approx0.03\,\mu
\mathrm{s}\,\mathrm{K}^{3}.$ Using $\frac{6}{\pi^{2}}v^{2}\tau=2\tau^{\ast}$
with
\begin{equation}
\tau^{\ast}=t/\tau_{\mathrm{e-ph}}\left(  T_{\Sigma}\right)  ,
\end{equation}
Eq.\thinspace(\ref{heat equation with time}) reduces to
\begin{equation}
\frac{\partial\theta^{2}}{\partial\tau^{\ast}}=-2\left(  \theta^{5}%
-\theta_{\mathrm{ph}}^{5}-f^{2}(\tau^{\ast})\right)  \label{largevlimit}%
\end{equation}
which for $\theta_{\mathrm{ph}}=f=0$ is simply equivalent to
\begin{equation}
\frac{\partial T}{\partial t}=-\frac{T}{\tau_{\mathrm{e-ph}}\left(  T\right)
}%
\end{equation}
expressing that the instantaneous decay rate of $T$ is exponential with a
characteristic time $\tau_{\mathrm{e-ph}}\left(  T\right)  .$
Equation\thinspace(\ref{largevlimit}) yields
\begin{equation}
\int\nolimits_{\theta^{2}(0)}^{\theta^{2}(\tau^{\ast})}\frac{\mathrm{d}%
w}{f^{2}(\tau^{\ast})+\theta_{\mathrm{ph}}^{5}-w^{5/2}}=2\,\tau^{\ast}.
\label{heatpho}%
\end{equation}
When $f(\tau)=1-H(\tau)$ and $T_{\mathrm{ph}}=0,$ the temperature decay from
$T_{\Sigma}$ has a simple form:%
\begin{equation}
\theta(\tau^{\ast})=\left(  1+3\tau^{\ast}\right)  ^{-1/3}. \label{coolph}%
\end{equation}
This temperature decay, which was directly measured in Ref.\thinspace
\onlinecite{Schmidt}  , follows a power law only, so that it takes a very long
time to recover the base temperature after the voltage is set to $0$, which is
due to the divergence of $\tau_{\mathrm{e-ph}}\left(  T\right)  $ when
$T\rightarrow0.$ The results of Eq.\thinspace(\ref{heatpho}) with $f=1$
(heating) and $f=0$ (cooling) are plotted in Fig.\thinspace\ref{E} in the case
$T_{\mathrm{ph}}=0$, with linear (top) and logarithmic (bottom) time scales.
The temperature rise is well approximated by $\theta(\tau^{\ast})\approx
\sqrt{2\,\tau^{\ast}}$ when $\tau^{\ast}\lesssim0.2$ (dashed line) and
$\theta(\tau^{\ast})\approx1-0.86\exp(-4.2\,\tau^{\ast})$ when $\tau^{\ast
}\gtrsim0.2$ (dotted line)$.$ More generally, when $\tau^{\ast}\ll1,$ for
$T_{\mathrm{ph}}\neq0,$
\begin{equation}
\theta(\tau^{\ast})\approx\sqrt{\theta^{2}(0)+2(1+\theta_{\mathrm{ph}}%
^{5}-\theta^{5}(0))\,\tau^{\ast}}. \label{approxheating}%
\end{equation}
\

Even though $v\gg1,$ we now estimate cooling by electron diffusion to the
connecting leads. Starting from a constant temperature $T_{0}$, cooling by
diffusion follows Eq.\thinspace(\ref{Tdexett}) with $a_{k}(\tau)=\left(
\frac{k_{B}T_{0}}{eV}\right)  ^{2}\frac{4}{\pi}\frac{e^{-\pi^{2}k^{2}\tau}}%
{k},$ and $T_{\mathrm{av}}/T_{0}\sim\exp(-5\tau).$ Because of this exponential
dependence, to be compared with the powerlaw (\ref{coolph}), diffusion can
contribute to the cooling when $t$ becomes comparable to $\tau_{D}$.

\bigskip

\subsection{Numerical application}

\begin{figure}[b]
\begin{center}
\includegraphics[width=1\columnwidth,clip]{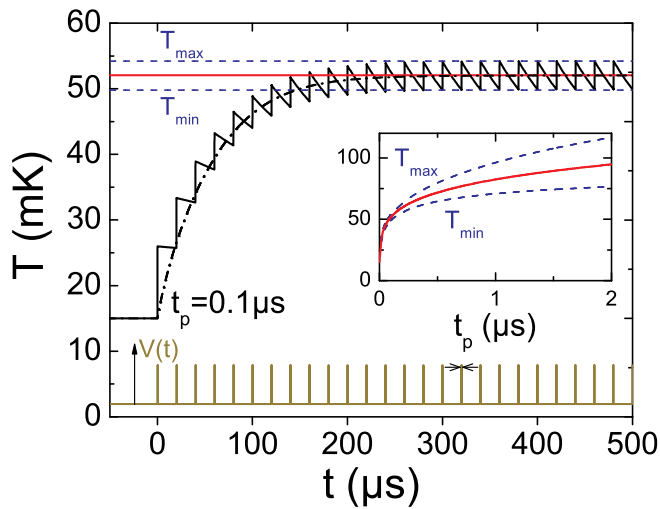}
\end{center}
\caption{(Color online) Main panel: Time-dependence of the temperature of a commercial macroscopic surface mount
500\thinspace$\Omega$ resistor (see text) heated by voltage pulses (bottom curve) of length $t_{p}=0.1\,\mu\mathrm{s}$
applied every $20\,\mu\mathrm{s.}$ Dashed-dotted line is the predicted heating with the average Joule power. Inset:
(blue dashed lines) minimal and maximal temperature reached in the stationary regime as a function of the pulse length
$t_{p}.$ In red solid line, temperature evaluated with the average Joule power.}%
\label{F}%
\end{figure}

In experiments where the voltage is applied in repeated pulses, heating is
reduced, and the temperature oscillates in time. To illustrate this effect, we
reconsider the second example of section IIC, but we now assume that the
voltage is applied during short pulses of duration $t_{p}=0.1\,\mu\mathrm{s},$
repeated every period $t_{r}=20\,\mu\mathrm{s}$ (which corresponds to actual
experimental conditions in Ref.\thinspace\onlinecite{Vion,theseAC}  ). We now
show that despite the short duty cycle $d=t_{p}/t_{r}=0.005$, heating is not
negligible.\ In our example, $v\sim30$, therefore the relevant timescale when
$V$ is applied is $\tau_{\mathrm{e-ph}}\left(  T_{\Sigma}\right)
=\gamma/\left(  \Sigma T_{\Sigma}^{3}\right)  \sim10\,\mu\mathrm{s}.$
Equation\thinspace(\ref{approxheating}) with $\theta_{\mathrm{ph}%
}=15\,\mathrm{mK}/\,T_{\Sigma}=0.1$ gives $T(t_{p})=0.17\,T_{\Sigma
}=25\,\mathrm{mK,}$ indicating slight heating by the first pulse. The
resistance then cools down during a time $t_{r}$ before the next pulse is
applied, following Eq.(\ref{heatpho}), to $T=0.169\,T_{\Sigma}$, hardly less
than at the end of the first pulse\cite{pasdiff}. The temperature rises
further during the next pulses, till steady oscillations establish. The full
time evolution of $T$ shown in Fig.\thinspace\ref{F} is obtained by iterating
Eq.(\ref{heatpho}). At each pulse, the temperature rise gets smaller than
during the preceeding pulse, because the starting temperature is larger and
the heat transfer to phonons becomes more efficient.\ For the same reason, the
cooling between the pulses gets more and more efficient. At $t\gtrsim
250\,\mu\mathrm{s,}$ a stationary regime is reached, with the reduced
temperature oscillating between $\theta_{\min}$ to $\theta_{\max}$ such that
$\int\nolimits_{\theta_{\min}^{2}}^{\theta_{\max}^{2}}\mathrm{d}%
w/(1+\theta_{\mathrm{ph}}^{5}-w^{5/2})=2\tau_{p}^{\ast}$ and $\int
\nolimits_{\theta_{\max}^{2}}^{\theta_{\min}^{2}}\mathrm{d}w/(\theta
_{\mathrm{ph}}^{5}-w^{5/2})=2(\tau_{r}^{\ast}-\tau_{p}^{\ast}).$ One obtains
$\theta_{\min}=0.33$ ($T_{\min}=49\,\mathrm{mK}$) and $\theta_{\max}=0.36$
($T_{\max}=54\,\mathrm{mK}$). The amplitude of the oscillations $\Delta
\theta=\theta_{\max}-\theta_{\min}$ is therefore very small. However, it
increases with $t_{p},$ as shown in the inset of Fig.\thinspace\ref{F}, and
can become sizeable.

The main features of the time evolution of the temperature can be calculated
more simply, from the average Joule power $d\times V^{2}/R.$ Using section II,
the characteristic temperature is then $T_{\Sigma}^{\mathrm{eff}%
}=52\,\mathrm{mK}$, which fits with the average temperature in the stationary
regime of the pulse sequence. According to section IIIB, this temperature is
reached in a time $\tau_{\mathrm{e-ph}}(T_{\Sigma}^{\mathrm{eff}})\sim
220\,\mu\mathrm{s.}$ The rise of temperature with time calculated with
Eq.\thinspace(\ref{heatpho}), in which all quantities ($\theta,$ $\tau^{\ast}%
$) are calculated using $T_{\Sigma}^{\mathrm{eff}},$ is shown as a dotted line
in Fig.\thinspace\ref{F}, and reproduces well the overall behavior. The
amplitude $\Delta\theta$ of the temperature oscillations can be evaluated
using Eq.\thinspace(\ref{approxheating}) under the assumption that the
starting temperature is $T_{\Sigma}^{\mathrm{eff}}$, which is a good
approximation for oscillations of small amplitude, and considering that the
voltage $V$ is always present during the pulse of duration $t_{p}$,
\textit{i.e.} with $\theta(0)=T_{\Sigma}^{\mathrm{eff}}/T_{\Sigma}=d^{1/5}$
and $\tau^{\ast}=t_{p}/\tau_{\mathrm{e-ph}}\left(  T_{\Sigma}\right)  .$ If
$\theta_{\mathrm{ph}}\ll1$ and $d\ll1,$ one obtains $\Delta\theta\approx
\theta(0)\tau^{\ast},$ an approximation only 20\% larger than the exact result
in the worst case of the inset of Fig.\thinspace\ref{F} ($t_{p}=2\,\mu
\mathrm{s).}$

\section{Summary}

The solution of the heat equation in a resistor is determined by a
characteristic temperature $T_{\Sigma}=\left(  V^{2}/\Sigma\Omega R\right)
^{1/5}.$ If $v=eV/k_{B}T_{\Sigma}\lesssim1,$ cooling by phonons is negligible
and $T(x)$ is given by Eq.\thinspace(\ref{hot}), the average temperature by
$T_{\mathrm{av}}=\sqrt{3}/8\,(eV/k_{B}).$ If $v\gtrsim10,$ the temperature is
$(T_{\Sigma}^{5}+T_{\mathrm{ph}}^{5})^{1/5}$ except at distances shorter than
$\sim5L/v$ from the ends. At $v\gtrsim4$ and $T_{\mathrm{ph}}=0,$ the average
temperature is $T_{\mathrm{av}}\approx T_{\Sigma}(1-1.16/v).$ Using these
results, we have calculated the decay of the Fano factors $F_{2}$ and $F_{3}$
relative to the second and third cumulants of current fluctuations with the
resistor length $L$. We have also addressed time-dependent situations to
describe the heating and cooling of resistors. If $v\lesssim1,$ the
characteristic timescale is the diffusion time $\tau_{D}$ and heating follows
$k_{B}T_{\mathrm{av}}(t)\simeq\frac{\sqrt{3}}{8}eV\sqrt{1-\exp(-10\tau)},$
cooling $k_{B}T_{\mathrm{av}}(t)\simeq\frac{\sqrt{3}}{8}eV\exp(-5\tau),$ with
$\tau=t/\tau_{D}.$ If $v\gg1,$ the instantaneous relaxation time is
$\tau_{\mathrm{e-ph}}\left(  T\right)  =\gamma/\left(  \Sigma T^{3}\right)  .$
Heating from $T(0)$ to $T_{\Sigma}$ is achieved in a time $\tau_{\mathrm{e-ph}%
}\left(  T_{\Sigma}\right)  ,$ following $\theta(\tau^{\ast})\approx
\sqrt{\theta(0)^{2}+2(1+\theta_{\mathrm{ph}}^{5}-\theta(0)^{5})\,\tau^{\ast}}$
at short times and $\theta(\tau^{\ast})\approx1-0.86\exp(-4.2\,\tau^{\ast})$
at long times, with $\theta=T/T_{\Sigma}$ and $\tau^{\ast}=t/\tau
_{\mathrm{e-ph}}\left(  T_{\Sigma}\right)  $. Cooling from a temperature
$T_{0}$ occurs very slowly, along a powerlaw: at $T_{\mathrm{ph}}=0,$
$T(t)/T_{0}=\left(  1+3t/\tau_{\mathrm{e-ph}}(T_{0})\right)  ^{-1/3}$.

Finally, we recall that the actual temperature can be higher than the
predictions made here for at least two reasons. First, the electronic
temperature can be larger than $T_{\mathrm{ph}}$ in the connecting wires
because of their finite resistivity\cite{Henny,NagaevN} or because of
imperfect thermalization to the cryogenic unit. Second, we have neglected the
Kapitza resistance\cite{Swartz}, due to which the phonon temperature inside
the resistor can differ from the bath temperature $T_{\mathrm{ph}}$. However,
this latest effect is relatively less important in very thin resistors because
the ratio of the heat flow from electrons to resistor phonons to the heat flow
from resistor phonons to substrate is proportional to the film
thickness\cite{Wellstood}.

\section{Appendix: calculation of the third cumulants of current in presence
of electron-phonon scattering}

The calculation of the third cumulant of current in presence of
electron-phonon scattering is an extension of the expressions of Pilgram et
al. \cite{Pilgram}. The third cumulant at zero frequency is expressed as a
function of the correlator between temperature and current fluctuations:%
\begin{equation}
S_{3}=\frac{6k_{B}}{R}\int\limits_{0}^{1}dx\langle\delta T(x)\delta I\rangle.
\label{S3ff}%
\end{equation}
To calculate the integrand, we start from the stochastic diffusion equation
for the fluctuations $\delta f$ of the electron energy distribution function
\begin{align}
&  \left(  \frac{\partial}{\partial t}-\frac{1}{\tau_{D}}\frac{\partial^{2}%
}{\partial x^{2}}\right)  \delta f-\delta I_{ee}-\delta I_{e-ph}\nonumber\\
&  =-e\delta\dot{\phi}\frac{\partial f}{\partial\varepsilon}-\frac{1}%
{L}\,\frac{\partial}{\partial x}\delta{F}^{imp}-\delta F^{ee} \label{26'}%
\end{align}
with $\delta I_{ee}$ the linearized electron-electron diffusion integral,
$\delta I_{e-ph}$ the linearized electron-phonon diffusion integral, $\delta
F^{imp}$ and $\delta F^{ee}$ random extraneous sources associated with
electron-impurity and electron-electron scattering. The correlation function
of extraneous sources is
\begin{align}
\langle\delta F^{imp}(\varepsilon,x)\delta F^{imp}(\varepsilon^{\prime
},x^{\prime})\rangle_{\omega}  &  =2\frac{D}{\nu_{F}\Omega}\,\delta
(x-x^{\prime})\delta(\varepsilon-\varepsilon^{\prime})\nonumber\\
&  \times f(\varepsilon,x)[1-f(\varepsilon,x)]. \label{<dF^2>}%
\end{align}
The energy distribution function is assumed to have a Fermi shape with
coordinate dependent temperature $T(x)$ and electrical potential $\phi(x):$
\begin{equation}
f(\varepsilon,x)=\left[  1+\exp\left(  \frac{\varepsilon-e\phi(x)}{k_{B}%
T(x)}\right)  \right]  ^{-1}.
\end{equation}
To derive the correlator $\langle\delta T_{e}(x)\delta I\rangle_{\omega}$,
Eq.\thinspace(\ref{26'}) is multiplied by $\varepsilon$ and integrated over
energy\cite{NagaevN}, assuming that the rate of energy dissipation associated
with electron--phonon scattering is of the form
\begin{equation}
\nu_{F}\int d\varepsilon\,\varepsilon I_{e-ph}=\Sigma\left[  T^{5}%
(x)-T_{ph}^{5}\right]  .
\end{equation}
The electron-electron collision integral and the associated extraneous source
drop out because of energy conservation, and one obtains
\begin{align}
&  \left(  \frac{\partial}{\partial t}-\frac{1}{\tau_{D}}\frac{\partial^{2}%
}{\partial x^{2}}\right)  \left(  e^{2}L_{o}\,T\delta T\right)  +5\nu_{F}%
^{-1}\Sigma\,T^{4}\,\delta T-\nonumber\\
&  -\frac{1}{\tau_{D}}\frac{\partial^{2}}{\partial x^{2}}\left(  e^{2}%
\phi\delta\phi\right)  =-\frac{1}{L}\int d\varepsilon\,\varepsilon
\frac{\partial}{\partial x}\delta{F}^{imp}. \label{26}%
\end{align}
Now we multiply Eq.\thinspace(\ref{26}) and the equation for the fluctuations
of the total current, which in the low-frequency limit reads\cite{Nagaev}
\begin{equation}
\delta I=\frac{e\nu_{F}\Omega}{L}\int d\varepsilon\int dx\,\delta F^{imp}.
\label{dI}%
\end{equation}
Upon averaging, it gives in the low-frequency limit
\begin{align}
&  \frac{\partial^{2}}{\partial x^{2}}\left[  L_{o}T\langle\delta T(x)\delta
I\rangle_{\omega}\right]  -L_{o}\alpha\,T^{4}\,\langle\delta T(x)\delta
I\rangle_{\omega}\nonumber\\
&  =-\frac{\partial^{2}}{\partial x^{2}}\left[  \phi\langle\delta\phi(x)\delta
I\rangle_{\omega}\right]  +\frac{2}{e}\,\frac{\partial}{\partial x}\int
d\varepsilon\,\varepsilon f(1-f). \label{27'}%
\end{align}
with $\alpha=5\Sigma\Omega R/L_{o}$. The right-hand side of this equation was
calculated in Ref.\thinspace\onlinecite{Pilgram}  . The solution of this
equation may be written in a symbolic form as
\begin{align}
\langle\delta T(x)\delta I\rangle_{\omega}  &  =\frac{2k_{B}}{L_{o}T}\,\left(
\frac{\partial^{2}}{\partial x^{2}}-\alpha\,T^{3}\right)  ^{-1}\times
\nonumber\\
&  \times\left\{  \frac{\partial(\phi T)}{\partial x}-\frac{\partial^{2}%
}{\partial x^{2}}\left[  \phi\left(  \frac{\partial^{2}}{\partial x^{2}%
}\right)  ^{-1}\frac{\partial T}{\partial x}\right]  \right\}  , \label{fluct}%
\end{align}
where the symbol $(\partial^{2}/\partial x^{2}-f)^{-1}$ is the Green's
function $G(x,y)$ such that $(\partial^{2}/\partial x^{2}-f)G(x,y)=\delta
(x-y)$ and $G(0,y)=G(x,0)=G(1,y)=G(x,1)=0.$ Using $\phi=-Vx,$ the expression
in brackets greatly simplifies:%
\begin{equation}
\left\{  \frac{\partial(\phi T)}{\partial x}-\frac{\partial^{2}}{\partial
x^{2}}\left[  \phi\left(  \frac{\partial^{2}}{\partial x^{2}}\right)
^{-1}\frac{\partial T}{\partial x}\right]  \right\}  =V(T-2T_{\mathrm{av}}).
\end{equation}
To calculate the third cumulant of the current, one has to solve
Eq.\thinspace(\ref{fluct}) and substitute the solution into Eq.\thinspace
(\ref{S3ff}). The generalized Fano factor $F_{3}=S_{3}/e^{2}I$ is then
\begin{align}
\bigskip F_{3}  &  =\frac{36}{\pi^{2}}\int\nolimits_{0}^{1}\mathrm{d}%
x\,\frac{1}{T(x)}\,\left(  \frac{\partial^{2}}{\partial x^{2}}-\alpha
\,T^{3}(x)\right)  ^{-1}\left\{  T-2T_{\mathrm{av}}\right\} \\
&  =\frac{36}{\pi^{2}}\int\nolimits_{0}^{1}\mathrm{d}x\,\mathrm{d}%
y\frac{1}{\theta(x)}\,G_{1}(\theta,x,y)\left\{  \theta(y)-2\theta
_{{\mathrm{av}}}\right\}
\end{align}
with $G_{1}(\theta,x,y)$ the Green's function such that
\begin{equation}
\left(  \frac{\partial^{2}}{\partial x^{2}}-\frac{15}{\pi^{2}}\,v^{2}%
\,\theta^{3}(x)\right)  G_{1}(\theta,x,y)=\delta(x-y),
\end{equation}
which can be calculated from\cite{Gutman}
\begin{equation}
G_{0}(x,y)=\left(  \frac{\partial^{2}}{\partial x^{2}}\right)  ^{-1}%
=\min(x,y)\,(\max(x,y)-1)
\end{equation}
using
\begin{equation}
G_{1}=\left(  1-\frac{15}{\pi^{2}}\,v^{2}\,G_{0}\theta^{3}(x)\right)
^{-1}G_{0}.
\end{equation}
In practice, we performed this calculation by discretization of the
coordinates and matrix inversion: the resistor is cut into $N$ pieces of
length $\varepsilon=1/N,$ and the function $G_{0}(x,y)$ is represented with a
matrix $G^{0}$ such that
\begin{equation}
G_{ij}^{0}=-\frac{\varepsilon}{N}\min(i,j)(\max(i,j)-1)
\end{equation}
$(0\leq i,j\leq N).$ The term $\frac{15}{\pi^{2}}\,v^{2}\,G_{0}\theta^{3}(x)$
is represented by the matrix $F$ build on the calculated temperature profile
$\theta(x)$ using
\begin{equation}
F_{ij}=G_{ij}^{0}\times\frac{15}{\pi^{2}}\,v^{2}\,\theta^{3}(j\,\varepsilon).
\end{equation}
We then invert the matrix $A$ with $A_{ij}=\frac{1}{\varepsilon}\delta
_{ij}-F_{ij}$ and compute $G^{1}=A^{-1}.G^{0}.$ Finally,
\begin{equation}
F_{3}=\frac{36}{\pi^{2}}\varepsilon\sum_{i,j}\frac{G_{ij}^{1}\{-\theta
(j\,\varepsilon)+2\theta_{\mathrm{av}}\}}{\theta(i)}.
\end{equation}

\begin{acknowledgments}
F. Pierre, Norman O. Birge and A.\ Anthore were involved in the early stages
of this work. Discussions with B. Reulet, H.\ Grabert, Yu. Gefen and within
the Quantronics group are gratefully acknowledged. We particularly appreciated
guidance from Hermann Grabert in the implementation of the calculation of the
third cumulant of noise. This work was partly funded by the Agence Nationale
de la Recherche under contract ANR-05-NANO-039.
\end{acknowledgments}


\begin{thebibliography}{9}
\bibitem {errorS}We have corrected the erroneous prefactor in Ref.\thinspace
\onlinecite{Steinbach}  .
\bibitem {notetaueph}Using a microscopic description of electron-phonon
scattering, $\gamma/\Sigma$ can be related to Fermi liquid parameters and to
the sound velocity\cite{Wellstood}. However, in experiments, a large
discrepancy was found with the corresponding theoretical
values\cite{Wellstood,theseFP,Qu}.
\bibitem {factor}We don't retain a factor 5 introduced in the definition of
$\tau_{\mathrm{e-ph}}$ in Ref.\thinspace\onlinecite{Schmidt}  . The length
$L_{\mathrm{e-ph}}$ defined in section IIA is related to $\tau_{\mathrm{e-ph}%
}(T)$ by $L_{\mathrm{e-ph}}=8^{5/2}3^{-3/4}\pi^{-1}\left(  D\,\tau
_{\mathrm{e-ph}}(eV/k_{B})\right)  ^{1/2}.$
\bibitem {pasdiff}Cooling by diffusion is here ineffective because of the too
long diffusion time $\tau_{D}\sim2\,\mathrm{ms}$.
\end{thebibliography}
\end{document}